\documentclass[twocolumn,prd,aps,groupedaddress,showpacs,
nofootinbib,preprintnumbers]{revtex4}

\usepackage{amsmath}
\usepackage{amsfonts}
\usepackage{graphicx}
\usepackage{bm}

\usepackage{varioref}
\usepackage{amssymb}

\begin{document}

\title{Stability of multi-field cosmological solutions in the presence
of a fluid}
\author{Jonathan Frazer}
\author{Andrew R. Liddle} 
\affiliation{Astronomy Centre, University of Sussex, Brighton BN1 9QH,
United Kingdom}
\date{\today}

\begin{abstract}
We explore the stability properties of multi-field solutions in the
presence of a perfect fluid, as appropriate to assisted quintessence
scenarios. We show that the stability condition for multiple fields $ \phi_i$ in identical potentials $V_i$ is simply $d^2V_i/d \phi_i^2 > 0$, exactly as in the absence of a fluid. A possible new instability associated with the fluid is shown not to arise in situations of cosmological interest.
\end{abstract}


\maketitle


\section{Introduction}

In models with multiple scalar fields, those fields may act
collectively to drive an accelerated expansion even if each field
individually were unable to, a phenomenon known as assisted inflation
\cite{LMS}. This possibility has now been widely explored for early
Universe inflation, and has begun to be considered for quintessence
models of the present acceleration as well \cite{CV,BP,KLT,Hetal,Tsujikawa,vBW}. In
the simplest scenario, a number of fields with identical uncoupled
potentials may be invoked.

Such scenarios clearly allow for solutions where the fields evolve
together, by symmetry. This is not sufficient however to demonstrate
assisted behaviour, as one must also check that such solutions are
stable. In cosmologies where the fields are the only components, this
was proven for the original exponential potential case in
Refs.~\cite{LMS,MW}, and Ref.~\cite{GA} subsequently provided the
general condition for stability which is simply that the potential for
the fields be convex, $d^2V_i/d \phi_i^2 >0$ where $ \phi_i$ are
the fields and each $V_i( \phi_i)$ has the same functional form.

In this article, we extend the result of Ref.~\cite{GA} to
allow for the presence of a fluid, hence obtaining the stability
conditions appropriate for assisted quintessence scenarios.

\section{Background evolution}

We consider a flat Friedmann--Robertson--Walker background with $n$
scalar fields and a fluid. The equations of motion are then the
Friedmann equation, fluid equation, and $n$ Klein--Gordon equations
\begin{eqnarray}
3H^2 = \rho + W +
\frac{1}{2}\sum_{i=0}^{n-1}\dot{ \phi_{i}}^2 \label{eq:friedmann}\,;\\ 
\dot{\rho} + 3H\gamma\rho = 0 \,; \label{eq:fluid}\\
\ddot{ \phi}_{i} + 3H\dot{ \phi} + \partial_{i}W = 0\,, \label{eq:KG}
\end{eqnarray}
where reduced Planck units $M_{\rm{Pl}} = c = 1$ have been used, $H$
is the Hubble parameter, $W \equiv
W( \phi_0, \phi_1..., \phi_{n-1})$ is the potential with both
self-interaction and interaction terms, $\gamma \equiv 1+ p/\rho$ is a
constant giving the equation of state of the fluid, and dots represent
derivatives with respect to synchronous time $t$.

A powerful method of exploring the stability of such a model is to
move into a Hamilton--Jacobi type formalism \cite{SB,EG} in which the role of the
clock is invested in a matter field so that one can get rid of the
unphysical degree of freedom represented by shifts in the time
coordinate $t$. Here we choose one of the scalar fields $ \phi_0 =
 \phi$ as the new time coordinate and perturb only the other
\mbox{$n-1$} fields $ \phi_i$. The actual field chosen is arbitrary; the instability we are exploring is in the difference between fields and would be identified whichever is chosen as the reference field.
Primes will indicate derivatives with
respect to the field. The Klein--Gordon equation for $ \phi$ is then
a constraint determining the relation between different time
coordinates. To rewrite the equations of motion \eqref{eq:friedmann},
\eqref{eq:fluid} and \eqref{eq:KG} in this new formalism, first we
combine them to obtain
\begin{equation}\label{eq:phi}
\dot{ \phi} = \frac{1}{A}\left(\frac{\rho'}{3H}-2H'\right),
\end{equation}
where 
\begin{equation}\label{eq:A}
A \equiv 1+ \sum_{i=1}^{n-1} \phi_i'^2,
\end{equation}
$\rho' \equiv \dot{\rho}/\dot{ \phi}$, $H' = \dot{H}/\dot{ \phi}$
and here and in what follows $i$ runs from $1$ to $n-1$. Rewriting
Eq.~(\ref{eq:friedmann}) as
\begin{equation}
3H^2 = \rho + W + \frac{1}{2}\dot{ \phi}^2\left(1+
\sum_{i=1}^{n-1} \phi_{i}'^2\right), 
\end{equation}
then substituting in Eq.~\eqref{eq:phi}, we get our Hamilton--Jacobi equation
\begin{equation}\label{eq:HJ}
 \frac{1}{2}\left(\frac{\rho'}{3H}-2H'\right)^2 - 3AH^2 + A(W+\rho) = 0.
\end{equation}
Note that if we send $\rho,\rho'\rightarrow0$, we recover the Hamilton--Jacobi
equation used in Ref.~\cite{GA}.

In the next section we will be writing down an equation of the form
$\delta X' = M \delta X$ where $X \equiv (H, \rho,  \phi_i,
 \phi_i' )^T$ is a $2n$-vector, so we combine Eq.~\eqref{eq:phi} with
Eq.~\eqref{eq:fluid} to obtain
\begin{equation}\label{eq:rhoprime}
\rho'^2  + 9\gamma A H^2\rho - 6HH'\rho' = 0.
\end{equation}
Finally we use the fact that 
\begin{equation}
\partial_{t}^{2} = \partial_{t}(\dot{ \phi}\partial_{ \phi}) =
 -(3H\dot{ \phi} + W')\partial_{ \phi} +
 \dot{ \phi}^2\partial_{ \phi}^{2},
\end{equation}
to get $n-1$ equations of the form
\begin{equation}\label{eq:phidoubleprime}
\dot{ \phi}^2 \phi_{i}'' - W' \phi_{i}+\partial_{i}W = 0.
\end{equation}
Note that this equation as written is independent of $H$ and $\rho$,
though in practice those quantities will influence the evolution of
$\dot{ \phi}$.
 
\section{Stability}

As in Ref.~\cite{GA} we wish to discuss the classical stability of
solutions of the form
\begin{equation}\label{eq:soln}
 \phi_{i}(t) =  \phi(t) \quad i = 0,...,n-1,
\end{equation}
such that the fields evolve together. Solutions of this form are only possible
if
\begin{equation}
\partial_iW\big|_{ \phi_{j}= \phi}=V'( \phi) \quad \forall i,j,
\end{equation}
where primes represent derivatives with respect to $ \phi$ and $V$ is defined
by this equation. If the fields are mutually decoupled, then each can be
written with its own potential $V_{i}$ obeying
\begin{equation}
V_{i}(x)+\Lambda_{i} = V(x) \quad \forall \ i. 
\end{equation}
Here the $\Lambda_{i}$ are constants and can all be absorbed into some
$\Lambda = \sum_{i}\Lambda_{i}$ acting as a cosmological constant, so
that $W = \sum_{i}V_{i} + \Lambda$. We will consider $\Lambda$ to be negligibly small or zero. The co-evolving field solutions can also exist within a limited class of models with cross-couplings between fields; for further discussion on
this class of solutions and the types of potential that permit it,
see Ref.~\cite{GA}.

We are now in a position to apply linear perturbation theory to
analyse the stability around solutions of the form
Eq.~\eqref{eq:soln}. When perturbing Eqs.  \eqref{eq:HJ},
\eqref{eq:rhoprime} and \eqref{eq:phidoubleprime} around
Eq.~\eqref{eq:soln}, some simplifications arise since
$ \phi_{i}''=0$ and \mbox{$A=n$} on the background $(\delta A =
2\sum_{i}\delta \phi_{i}')$. Perturbing the Hamilton--Jacobi
equation we find
\begin{equation}
\delta H' = \mathrm{A}\delta H + \mathrm{B}\delta\rho + \mathrm{C}\delta \phi_i
+\mathrm{D}\delta \phi_{i}'\,,
\end{equation}
where
\begin{eqnarray}
        \mathrm{A} &=& \frac{1}{36 H^3 n \dot{ \phi }} \, \left[-24 H H' \rho '+36 H^2 n \left(-6 H^2+W+\rho \right) \right. \nonumber \\ &&\left. \qquad  \qquad +72 H^2 H'^2+\rho' \right]\,;
\end{eqnarray}
\begin{eqnarray}
        \mathrm{B} & = & \frac{-12 H H' \rho '+18 H^2 n \rho +\rho '}{36 H^2         n \rho  \dot{ \phi }}\,; \\
        \mathrm{C} & = &\frac{V'}{2 \dot{ \phi }}\,; \\
        \mathrm{D} & =& -\frac{\dot{ \phi }}{2 H}\,.
\end{eqnarray}
From perturbing Eq.~\eqref{eq:rhoprime} we obtain 
\begin{equation}
        \delta \rho' = \mathrm{E}\delta H + \mathrm{F}\delta\rho + \mathrm{G}\delta \phi_i
        +\mathrm{H}\delta \phi_{i}',
\end{equation}
where
\begin{eqnarray}
        \mathrm{E} &=&-\frac{3 \left(\mathrm{A} H \rho '+\rho  H'-3 H n \gamma  \rho         \right)}{3 HH'-\rho '} \,; \\
        \mathrm{F}&=&\frac{3 H \left(3 H n \gamma -2 \mathrm{B} \rho '\right)}{6         HH'-2\rho '} \,; \\
        \mathrm{G}&=&\frac{6 \mathrm{C} H \rho '}{2 \rho '-6 H H'} \,; \\
        \mathrm{H}&=&\frac{3 H \left(3 H \gamma  \rho -\mathrm{D} \rho '\right)}{3 H         H'-\rho '} \,.
\end{eqnarray}
(Note that we use Roman typeface to label coefficients and that in
particular $H$, the Hubble parameter, and $\mathrm{H}$, the
coefficient of $\delta \phi_{i}'$, should be distinguished.)

Finally, from perturbing Eq.~\eqref{eq:phidoubleprime} we obtain
\begin{equation}
\delta \phi_{i}'' = \mathrm{I}_i\delta \phi_{i}+\mathrm{J}\delta \phi_{i}',
\end{equation}
where 
\begin{equation}
\label{eq:I}
        \mathrm{I}_i=-\frac{1}{\dot{ \phi}} \sum_{j=1}^{n-1} \left(\partial_{i}\partial_{j}W-\partial_j W' \right) \quad ; \quad \mathrm{J}=\frac{V'}{\dot{ \phi}} \,.
\end{equation}
We thus have a $2n \times 2n$ matrix of the form
\begin{align}
\begin{pmatrix}
 \mathrm{A} & \mathrm{B} & \mathrm{C_{1}} &\mathrm{C_{2}}&\cdots&\mathrm{C_{n-1}} & \mathrm{D_{1}}&\mathrm{D_{2}}& \cdots & \mathrm{D_{n-1}} \\
 \mathrm{E} & \mathrm{F} & \mathrm{G_{1}} &\mathrm{G_{2}}&\cdots&\mathrm{G_{n-1}} & \mathrm{H_{1}}&\mathrm{H_{2}}& \cdots & \mathrm{H_{n-1}} \\
     0      &      0     &      0     &      0     &\cdots  & 0 & 1 & 0 &
     \cdots & 0   \\
      0      &      0     &      0     &      0     & & 0 & 0 & 1 &
      & 0   \\
     \vdots & \vdots& \vdots& &\ddots & \vdots & \vdots& & \ddots\\
      0 & 0 & 0 & 0 & & 0 & 0 & 0 &  & 1 \\
     0      &      0     & \mathrm{I_{1}} & 0 & \cdots &0 & \mathrm{J_{1}}
     & 0 & \cdots & 0 \\
     0  & 0 & 0 & \mathrm{I_{2}} & & 0 & 0 & \mathrm{J_{2}} &  & 0\\
     \vdots & \vdots & \vdots & &  \ddots &  & \vdots & &\ddots\\  
     0 & 0&0 &0 & &\mathrm{I_{n-1}} & 0 & 0 &  & \mathrm{J_{n-1}}\\
\end{pmatrix},
\end{align}
such that, suppressing the subscript $i$s where possible, the system is described by
\begin{align}
\begin{pmatrix}
  \delta H'  \\
  \delta \rho'  \\
  \delta  \phi_i' \\
  \delta  \phi_i''
\end{pmatrix}=
\begin{pmatrix}
 \mathrm{A} & \mathrm{B} & \mathrm{C} & \mathrm{D} \\
 \mathrm{E} & \mathrm{F} & \mathrm{G} & \mathrm{H} \\
     0      &      0     &      0     &      1     \\
     0      &      0     & \mathrm{I}_i & \mathrm{J} \\  
\end{pmatrix}
\begin{pmatrix}
 \delta H  \\
  \delta \rho  \\
  \delta  \phi_i \\
  \delta  \phi_i'
\end{pmatrix}.
\end{align}
To analyse the solutions we use the characteristic equation
$\left\vert M-\lambda I_{2n}\right\vert$=0 to find eigenvalues. We
then require that all eigenvalues satisfy $\Re(\lambda)<0$ for
stability.  By standard manipulation of the determinant, the
characteristic equation can be reduced to the block-diagonal form
\begin{align}
\label{e:bdce}
\begin{vmatrix}
\mathrm{A}-\lambda & \mathrm{B}           & 0 & 0 \\
\mathrm{E}         & \mathrm{F}-\lambda   & 0 & 0 \\
0                  & 0                    & -\lambda & 1 \\
0                  & 0                    &\mathrm{I}_i &\mathrm{J}-\lambda\\
\end{vmatrix}=0,
\end{align}
and so the terms C, D, G, and H do not influence stability.  This
expression demonstrates that perturbations in the scalar field are
independent of perturbations in the metric $\delta H$ and the fluid
$\delta\rho$. In this paper, we shall concentrate on the case where
the fields are decoupled such that $\partial_j \partial_i
W( \phi)=V''( \phi)\delta_{ij}$, $\partial_{j}W'=0$ and $j = i$ in
Eq.~\eqref{eq:I}, but since the scalar fields' perturbations are
unaffected by the presence of a fluid, the more complex models
discussed in Ref.~\cite{GA} are still valid here.

The eigenvalues are 
\begin{equation}\label{eq:lambdarho}
\lambda_{\delta H,\delta\rho} = \frac{1}{2} \left(\mathrm{A+F}\pm\sqrt{\mathrm{A}^2-2 \mathrm{A} \mathrm{F}+4 \mathrm{EB}
   +\mathrm{F}^2}\right)
\end{equation}
and
\begin{equation}\label{eq:lambdascalar}
\lambda_{\delta \phi_{i}}=\frac{1}{2} \left(\mathrm{J}\pm\sqrt{\mathrm{J}^2+4
   \mathrm{I}}\right).
\end{equation}
Eq.~\eqref{eq:lambdascalar} is a $(n-2)$-degenerate pair. We use the
convention $\dot{ \phi}>0$ and $V'<0$ as we are interested in
situations where the field is rolling down the
potential. Consequently, if we substitute Eq.~\eqref{eq:I} back into
Eq.~\eqref{eq:lambdascalar}, taking account of the simplification
above
\begin{equation}
\lambda_{\delta \phi_{i}}=\frac{V'}{2\dot{ \phi}^2}\left(1 \pm\sqrt{1-\frac{4\dot{ \phi}^2V''}{V'^2}}\right)
\end{equation}
we find that the requirement for stability is simply
\begin{equation}
V'' >0\,.
\end{equation}
This is intuitively the case as we know from being able to write the
determinant in block-diagonal form, that the scalar fields evolve
independently of the metric and thus evolve independently of each
other. So for the perturbations to die away we require that the fields
further down the potential are evolving more slowly than those further
up the potential, namely that the potential is convex.  This is unchanged
from the result of Ref.~\cite{GA}, i.e.\ the stability within the
scalar sector is unchanged by the presence of the fluid.

Analysing Eq.~\eqref{eq:lambdarho} is less straightforward. In the absence of
a fluid the requirement for stability is simply that found in Ref.~\cite{GA},
\begin{equation}
-\frac{3H}{\dot{ \phi}}<0,
\end{equation}
which is automatically satisfied.  This can be derived by noting that in this limit the product BE
vanishes, and F is irrelevant as there is no $\delta \rho$, and so the
RHS of Eq.~\eqref{eq:lambdarho} simply becomes $\mathrm{A}$ which can be
simplified to the above form. 
In the presence of the fluid, there is a new perturbation degree of freedom corresponding to a shift in the fluid density with respect to the value of the scalar field $ \phi$ acting as the clock, which could in principle yield a new instability. As stability requires that the 
real part of both eigenvalues is
negative, we need to check that both $\mathrm{A}+\mathrm{F} < 0$ and $\mathrm{BE-AF}<0$ are satisfied.

The condition for the former to be satisfied can be shown, after some algebra, to be
\begin{align}\label{A+F ineq}
n \dot{ \phi }^2 \left[36 H^3 n (\gamma +1) \dot{ \phi }^2+6 H \gamma ^2 \rho  \left(\rho -6 H^2\right)+\gamma  \rho  \dot{\phi }\right] \nonumber\\
 \lessgtr \gamma ^2 \rho  \left(6 H^2+\rho \right) \left(6 H \gamma  \rho +\dot{ \phi }\right),
\end{align}
provided
\begin{equation}\label{eq:rhophiineq}
\gamma\rho-n\dot{ \phi}^2\gtrless 0.
\end{equation}
For general $\gamma$ the condition for the latter is rather long, so
we show only the simpler case of matter domination
$(\gamma = 1)$:
\begin{equation}\label{eq:BE-AF ineq}
108 H^4 n^2 \dot{\phi }^3\lessgtr\rho  \left(6 H \rho +\dot{ \phi }\right) \left[n \dot{ \phi } \left(3 H+\dot{ \phi }\right)+\rho \right]
\end{equation}
provided
\begin{equation}\label{eq:subcond2}
\rho-n\dot{ \phi}^{2}\gtrless 0.
\end{equation}

Both conditions have a direction of inequality that depends on a subsidiary condition. We first note that these will not be satisfied in complete generality, because when the subsidiary condition is saturated the main condition is not, meaning there must be some regime where the main condition fails in the vicinity of the region where the subsidiary condition flips.

However, in situations of practical cosmological interest, Eqs.~(\ref{eq:rhophiineq}) and (\ref{eq:subcond2}) will both be satisfied with a positive inequality, indicating the fluid dominating over the combined field kinetic energy. This is of course true in the early stages of cosmological evolution where the fluid dominates the total density, but it remains true up to the present, even though the total field energy density dominates, as the observed equation of state $w\equiv \gamma-1$ close to $-1$ requires that the field kinetic energy is still subdominant to the fluid. Only sometime in the future is the field kinetic energy expected to overtake the fluid.

Qualitatively, taking the kinetic part
of the scalar field to be subdominant, $\dot{ \phi}^{2}\ll\rho+W$, in
Eq.~\eqref{A+F ineq} we require LHS $<$ RHS which is satisfied thanks to
the LHS being suppressed by a factor of
$\dot{ \phi}^{2}$. Similarly the LHS of Eq.~\eqref{eq:BE-AF ineq} is suppressed by a factor $\dot{ \phi}^3$.\footnote{We also note that were we in a kinetic-dominated case, reversing the required inequality, those suppressions would become enhancements suggesting that stability would be restored deep in the kinetic-dominated regime.}

We can make this argument quantitative by defining a parameter $\alpha$ that measures the dominance of the fluid:
\begin{equation}
\rho = 3H^2\left(1+\alpha\right) \quad; \quad\alpha \equiv -\left(\frac{n\dot{ \phi}^2}{6H^2}+\frac{W}{3H^2}\right)\,.
\end{equation}
Reintroducing Planck masses for clarity, then Eq.~\eqref{eq:BE-AF ineq} becomes
\begin{eqnarray}\label{eq:BE-AF alpha}
&& \hspace*{-0.5cm} 108 H^2 n^2 \dot{\phi }^3<162 H^5 M_{\rm Pl}^3 (\alpha +1)^3\nonumber \\
&& +162 H^4 M_{\rm Pl}^2 n (\alpha +1)^2 \dot{\phi } \nonumber \\
&& +54 H^3 M_{\rm Pl} n (\alpha +1)^2 \dot{\phi }^2+9 H^2 M_{\rm Pl}^4 (\alpha +1)^2 \dot{\phi}
\nonumber \\
   &&+9 H M_{\rm Pl}^3 n (\alpha +1) \dot{\phi }^2+3 M_{\rm Pl}^2 n (\alpha +1) \dot{\phi }^3\,,
\end{eqnarray}
where each term is positive.

The first situation of interest is
when we have matter domination with the scalar fields highly subdominant,
as it is interesting to know if we have already reached the
solution \eqref{eq:soln} by the time the Universe starts to accelerate. Since $\alpha \rightarrow 0$ in this situation and we have that $n\dot{ \phi}^2
\ll 6H^2 M_{\rm Pl}^2$, we see that even the first term on the RHS is sufficient to
guarantee the satisfaction of the inequality.

The second case we are interested in is our present one, in the vicinity of the onset of acceleration. Then the energy density of the scalar field and fluid are
of similar order. The condition for acceleration is
\begin{eqnarray}\label{eq:accellcond}
\ddot{a}>0 \Longleftrightarrow n\epsilon < 1-\frac{\gamma\rho}{2H^2},
\end{eqnarray}
where
\begin{eqnarray}
\epsilon \equiv \frac{\left(\frac{\rho'}{3H}-2H'\right)^2}{2nH^2}
\end{eqnarray}
is a slow-roll parameter. Note that for a single scalar field and no fluid, we recover the first Hamilton--Jacobi slow-roll parameter 
\begin{eqnarray}
\epsilon_{\rm{H}}\equiv2\left(\frac{H'}{H}\right)^2.
\end{eqnarray} 
We thus see that stability in this situation is a little more complicated.
If $\rho>H^2$ then stability is guaranteed but if $\rho<H^2$ then acceleration
can take place for either sign in Eq.~\eqref{eq:rhophiineq} and since equality
does not occur at the same time in any, let alone all the conditions \eqref{A+F ineq} \eqref{eq:rhophiineq} \eqref{eq:BE-AF ineq},  stability is no longer guaranteed. Having said that, we know 
at the present time the equation of state parameter $w\simeq-1$ and so we
still expect Eq.~\eqref{eq:rhophiineq} to be positive and thus, since $1+\alpha$
is still of order 1, Eq.~\eqref{eq:BE-AF alpha} is again satisfied. Using
the same approach one can show that Eq.~\eqref{A+F ineq} is also satisfied
under these conditions and hence we have stability both at early times and
during conditions similar to the present day. 

We end this section with a brief comment on the inclusion of a separate constant $\Lambda$ term. For the scalar field sector the situation is unchanged by the introduction of a fluid; Ref.~\cite{GA} showed that introduction of $\Lambda$ could change the type of stability (oscillatory or non-oscillatory about the background solution), but could not change stability to instability. However the new fluid degree of freedom obeys a more complex equation and it does not appear possible to make a general statement on the effect of $\Lambda$. Nevertheless, cosmologies with a fluid and fields are already capable of explaining observations without needing $\Lambda$ and in this set-up a zero value of $\Lambda$ is the most natural choice.

\section{Conclusions}

We have derived a set of stability conditions against classical perturbations for multi-field cosmological solutions in the presence
of a fluid. Our focus has been on the co-evolving (`assisted') class of solutions for which a general stability analysis has been possible; for more general cross-coupled potentials such solutions typically will not exist and any stability properties must be assessed on a case-by-case basis.
Within the scalar sector, the condition for stability is independent of the presence of the fluid, as seen immediately from the block-diagonal form of the characteristic equation, Eq.~(\ref{e:bdce}), indicating that field perturbations are decoupled from the fluid sector. 

The introduction of the fluid introduces a new degree of freedom, corresponding to a shift in the fluid density with respect to the scalar field `clock'. Stability under this perturbation does not appear to be guaranteed for arbitrary backgrounds, but we have shown that for situations of physical interest this stability condition is always met. 

Our stability criteria are local, rather than global, conditions on the solutions; since they depend on the conditions on the potential they may be satisfied during some parts of the evolution and not others. An example would be multiple fields on a cosine potential as in axion models, e.g.\ as in Ref.~\cite{KLS}, where the co-evolving solution would be unstable around the maximum ($V''<0$) and stable around the minimum ($V''>0$). In regimes where the co-evolving solution is unstable it would clearly be extremely fine-tuned, whereas in those where it is stable there will be a basin of attraction around the co-evolving solution making co-evolving evolution much more likely.

Our results also introduce a new perspective on tracking solutions \cite{SWZ,LS}. Normally one considers the fluid evolution $\rho(a)$ as the fixed quantity; the field $ \phi$ is then shown to have various possible behaviours as a function of $a$, which converge together if the tracking criteria are met. In our formalism, instead it is $ \phi_i( \phi)$ which is fixed to be equal to $ \phi$, and $\rho$ is perturbed as a function of $\phi$. It would be interesting to make a more extensive exploration of tracking properties in the Hamilton--Jacobi formulation.

\acknowledgments
The authors were supported by STFC (UK). 


\end{document}